\documentclass[useAMS,usenatbib]{mnras}
\usepackage{graphicx}
\usepackage{amsmath,amssymb}
\usepackage{enumitem}
\usepackage{placeins}
\usepackage[caption=false]{subfig}
\usepackage{multirow}
\usepackage{booktabs}

\def\Teff{$T_{\rm eff}$}
\def\logg{$\log\,g$}

\def\Vt{V${\rm t}$}

\def\Rg{$R_{G}$}
\def\Rsun{$R_{G,\odot}$}
\def\kms{km s$^{-1}$}

\def\FeH{$\mathrm{[Fe/H]}$}

\usepackage{color}

\title[Galactic abundance gradients]
{The MAGIC project. III. Radial and azimuthal Galactic abundance gradients using classical Cepheids}
\author[V.~Kovtyukh et al.]
{V.~Kovtyukh$^{1}$\thanks{Based on observations obtained with the
Southern African Large Telescope (SALT), programme \mbox{2016-1-MLT-002}
(PI: Kniazev).
\newline Tables~\ref{spectra}-\ref{abund3} are only available in electronic form
at the CDS via the anonymous ftp to cdsarc.u-strasbg.fr
(130.79.128.5)
or via http://cdsweb.u-strasbg.fr/cgi-bin/qcat?J/MNRAS/(vol)/(page)
\newline E-mail: vkovtyukh@ukr.net},
B.~Lemasle,$^{2}$
G.~Bono,$^{3,4}$
I.~A.~Usenko,$^{5}$
R.~da Silva,$^{4,6}$
A.~Kniazev,$^{7,8,9}$
\newauthor
E.~K.~Grebel,$^{2}$
I.~L.~Andronov,$^{10}$
L.~Shakun,$^{1}$
L.~Chinarova$^{10}$
\\
$^{1}$ Astronomical Observatory, Odessa National University, Shevchenko Park, 65014, Odessa, Ukraine\\
$^{2}$ Astronomisches Rechen-Institut, Zentrum f\"ur Astronomie der Universit\"at Heidelberg, M\"onchhofstr. 12-14, 69120 Heidelberg, Germany\\
$^{3}$ Dipartimento di Fisica, Universit\'a di Roma Tor Vergata, via della Ricerca Scientifica 1, 00133 Rome, Italy\\
$^{4}$ INAF -- Osservatorio Astronomico di Roma, via Frascati 33, Monte Porzio Catone, Rome, Italy\\
$^{5}$ Mykolaiv Astronomical Observatory Research Institute, Obsevatorna 1, Mykolaiv 54030, Ukraine\\
$^{6}$ Agenzia Spaziale Italiana, Space Science Data Center, via del Politecnico snc, 00133 Rome, Italy\\
$^{7}$ South African Astronomical Observatory, PO Box 9, 7935 Observatory, Cape Town, South Africa \\
$^{8}$ Southern African Large Telescope Foundation, PO Box 9, 7935 Observatory, Cape Town, South Africa \\
$^{9}$ Sternberg Astronomical Institute, Lomonosov Moscow State University, Universitetskij Pr. 13, Moscow 119992, Russia\\
$^{10}$ Odessa National Maritime University, Mechnikova St. 34, Odessa 65029, Ukraine\\
}

\date{Accepted 2021 November 29. Received 2021 November 24; in original form 2021 September 16}

\pubyear{2021}

\begin{document}
\label{firstpage}
\pagerange{\pageref{firstpage}--\pageref{lastpage}}
\maketitle

\begin{abstract}
\par Radial abundance gradients provide sound constraints for chemo-dynamical models of galaxies. Azimuthal variations of abundance ratios are solid diagnostics to understand their chemical enrichment. In this paper we investigate azimuthal variations of abundances in the Milky Way using Cepheids. We provide the detailed chemical composition (25 elements) of 105 Classical Cepheids from high-resolution SALT spectra observed by the MAGIC project. Negative abundance gradients, with abundances decreasing from the inner to the outer disc, have been reported both in the Milky Way and in external galaxies, and our results are in full agreement with literature results. We find azimuthal variations of the oxygen abundance [O/H]. While a large number of external spirals show negligible azimuthal variations, the Milky Way seems to be one of the few galaxies with noticeable [O/H] azimuthal asymmetries. They reach $\approx$0.2~dex in the inner Galaxy and in the outer disc, where they are the largest, thus supporting similar findings for nearby spiral galaxies as well as recent 2D chemo-dynamical models. 
\end{abstract}
\begin{keywords}
stars: abundances -- stars: variables: Cepheids --
Galaxy:
\end{keywords}

\section{Introduction}

\par Radial abundance gradients provide sound constraints to galaxy formation scenarios. Indeed, the star formation history, the accretion history, the radial migration of stars and the radial flows of gas, and their variations with the Galactocentric distance, simultaneously determine the shape of abundance gradients. Chemo-dynamical models of the Milky Way must therefore reproduce the observed gradients and their evolution with time. 
\par The advent of Integral Field Spectrographs (IFS) has provided the opportunity to explore a large number of nearby spiral galaxies \citep[for a review, see][]{Sanchez2020}, most of them showing negative abundance gradients from the inner to the outer disc and a flattening of the gradient in the outermost regions \citep{Sanchez-Menguiano2018}. Spectroscopic investigations covering the outermost regions of large spiral galaxies also show a well-defined 
negative age gradient \citep[e.g.,][]{Li2015}, thus supporting the "inside-out" scenario \citep[e.g.,][]{Matteucci1989,Spitoni2011}.  

\par In the Milky Way, the task is more complicated:
\begin{itemize}[nosep]
    \item[-] only recently, Gaia \citep[][]{Gaia2018} provided accurate parallax-based distances for a large number of tracers (in particular, red giant branch (RGB) stars), but they are limited to the extended ($\approx$5~kpc) Solar neighborhood, and RGB ages could be determined only for a smaller, nearby sample; 
    \item[-] tracers with both accurate distances (even far from the Sun) and ages, such as Cepheids and open clusters, do not cover the entire age range, they are rarer than RGBs, and samples are not yet complete, especially in the outer disc;
    \item[-] extinction hampers the detection and the spectroscopic analysis in the optical regime of (all) distant tracers.
\end{itemize}
As a result, the radial distribution of oxygen and iron abundances (among others) in the Galactic disc is still not clear. Using Cepheids, linear [Fe/H] gradients with a slope of roughly --0.05~dex kpc$^{-1}$ describe well the data over distances in the [5--15] kpc range \citep[e.g.,][]{Kovtyukh2005,Genovali2014,Luck2018}, but more complicated features including a flattening in the outer disk and a sharp variation  of $\approx$0.2 dex in the slope have also been suggested \citep[e.g., ][]{Luck2003}. Similar radial gradients and a flattening for Galactocentric distances 
larger than 14/16 kpc have also been found by \citet[][]{Donor2020} and by \citet[][]{Magrini2017} using open clusters covering a moderate range in ages. A sharp variation in the slope of the radial gradient across the solar circle was also suggested 
by \citet[][]{Twarog1997} by using a sizable sample of open clusters.
Pure 1D chemical evolution models \citep[e.g., ][]{Cescutti2007} as well as hybrid models, which include in addition 3D kinematics from numerical simulations \citep[e.g., ][]{Minchev2014,Kubryk2015} reproduce quite well the Cepheid and open cluster observations.
We note in passing that, given their very young age ($<$300\,Myr), Cepheids are presumably only marginally affected by radial migration.

\par Significant azimuthal variations in [O/H] have for now been reported in only a few galaxies \citep[][]{Li2013,Sanchez2015,Sanchez-Menguiano2016,Zinchenko2016,Ho2017,Vogt2017,Ho2018}. However, we still lack solid empirical/theoretical evidence to assess whether they are either local anomalies or the aftermath of secular evolutions. In the Milky Way, various authors have reported large scale inhomogeneities in the azimuthal gradient, for instance \citet[][]{Davies2009} using open clusters, \citet[][]{Luck2006,Lemasle2008,Genovali2014} using Cepheids, \citet[][]{Balser2011,Balser2015} using H~{\sc ii} regions. The latter studies found azimuthal abundance variations for [O/H], leading to different slopes (by a factor about two) for the [O/H] Galactocentric radial gradient, depending on the Galactic azimuth range sampled, and those results were confirmed by \citet[][]{Wenger2019}. By using a 2D chemo-dynamical model taking account for the Milky Way spiral pattern
\citet[][]{Spitoni2019} found a scatter in the azimuthal variations of the oxygen abundance gradient and suggested that it is the consequence of the multiple spiral modes moving at different rotational velocities.

\par In this paper, we present new results of the MAGIC project (\underline{\bf M}ilky W\underline{\bf A}y \underline{\bf G}alaxy w\underline{\bf I}th SALT spe\underline{\bf C}troscopy; \cite{Kniazev2019}, a large spectroscopic survey that uses spectral instrumentation of the Southern African Large Telescope \citep[SALT, ][]{Buck2006,Dono2006}. MAGIC targets pulsating variable stars, in particular Cepheids in order to study the Milky Way chemical evolution. We provide the stellar parameters of 105 Classical Cepheids and the abundances of 25 elements, based on 122 high-resolution SALT spectra. Our spectroscopic analysis is described in Sect.~\ref{data+analysis}, and the results in terms of radial gradients and azimuthal inhomogeneities are presented in Sect.~\ref{results}

\section{Spectroscopic analysis}
\label{data+analysis}

\subsection{Data}
\label{data}

\par We have used the SALT HRS high resolution spectrograph operated in the medium-resolution mode (MR: R$\sim$37\,000--39\,000). It provides two spectra in the blue and red arms over a spectral range of $\approx$3900--8900~\AA. Both the blue and red CCDs were read out in the standard 1$\times$1 binning mode. Three spectral flats and one spectrum of ThAr+Ar lamps were obtained in that mode from a one-week series of HRS calibrations, which yield an accuracy of the velocity measurements of about 150 m/s in the MR mode \footnote{See \href{http://pysalt.salt.ac.za/proposal\_calls/current/ProposalCall.html\#h.oomdn5s2jnj6}{this link} for more details}.
The HRS primary data reduction, including overscan and gain corrections, as well as bias subtractions, was done through the SALT science pipeline \citep{Cr10}. The spectroscopic reduction of the HRS data was performed using our own HRS pipeline, described in detail in
\cite{KGB16} and \cite{Kniazev2019}. Relevant information about the spectra is given in Table~\ref{spectra}.

\subsection{Determination of abundances}
\label{analysis}

\par The effective temperature, \Teff, was derived from line-depth ratios \citep{Kov2007}, a technique commonly employed in studies of Cepheid variables \citep[e.g.,][]{Andrievsky2002a,Lemasle2007,Kovtyukh2016,Luck2018,Proxauf2018}. Once \Teff\ determined,
the surface gravity \logg\ was computed by imposing the iron ionisation balance. The microturbulent velocity, \Vt, was derived assuming that there is no correlation between the iron abundance A(Fe), obtained from Fe~{\sc i} lines, and the equivalent widths (EW) of the same lines. The adopted value for [Fe/H] is the one derived from the Fe~{\sc i} lines, since we imposed the ionisation balance and because they outnumber Fe~{\sc ii} lines. The atmospheric parameters \Teff, \logg\ and \Vt\ are listed in Table~\ref{spectra}.

\begin{table*}
\begin{center}
\caption[]{Information on the spectra and atmospheric parameters for the investigated Cepheids. The modified Julian date (MJD) of the observations and the exposure times are given in columns (2) and (3). The following columns display \Teff, its uncertainty ($\sigma$), the number N of line depth ratios used to derive \Teff, and the quantity $\sigma/\sqrt{N}$. Then the values of \logg\ and \Vt\ are provided, followed by [Fe/H], its uncertainty and the number of Fe\,I lines used. Additional remarks are provided when necessary. Only the first ten lines of the table are shown. The full table is available in electronic form as Table~1.}
\label{spectra}
\begin{tabular}{rrrrrrrrrrrrl}
\hline
   Object      &    MJD       & exp. time &  \Teff & $\sigma$ & N &$\sigma/\sqrt{N}$&  \logg & \Vt   & \FeH   & sigma &   N   & Remarks\\
               &  2450000+    & (s)     &    (K) &  (K)  &   &   (K)   &        & \kms  &  (dex) &      &       &        \\
\hline
   RW CMa      &     7733.40182 &  1597 &   6019 & 158&  69 &     19.1 &  1.60  &  2.90 &   0.07 & 0.07 &  149  &  \\
   RW CMa      &     7733.41557 &  1597 &   5992 &  81&  57 &     10.7 &  1.60  &  3.20 &   0.03 & 0.06 &  137  &  \\
   RW CMa      &     7755.34911 &  1597 &   5993 & 104&  60 &     13.4 &  1.70  &  2.90 &   0.05 & 0.10 &   99  &  \\
 V384 CMa      &     7763.31139 &  3000 &   5554 & 146&  59 &     18.9 &  1.65  &  2.30 &   0.02 & 0.06 &  114  &  \\
   CC Car      &     7530.33756 &  3000 &   5726 &  91&  78 &     10.3 &  1.70  &  2.90 &   0.18 & 0.10 &  196  &  \\
   CR Car      &     7736.51354 &  2248 &   5859 &  87&  71 &     10.4 &  1.50  &  2.80 &   0.14 & 0.07 &  194  &  \\
   CT Car      &     7531.30744 &  3340 &   5298 &  97&  67 &     11.8 &  1.00  &  3.20 &   0.07 & 0.06 &  195  &  \\
   FN Car      &     7737.55065 &  2212 &   5784 &  69&  72 &      8.2 &  1.70  &  2.60 &   0.15 & 0.07 &  198  &  \\
   FQ Car      &     7736.54721 &  2672 &   5210 &  57&  69 &      6.9 &  1.30  &  3.10 &   0.17 & 0.08 &  183  &  \\
   FZ Car      &     7548.29140 &  2943 &   5970 &  87&  76 &     10.0 &  1.90  &  2.60 &   0.13 & 0.12 &  197  &  \\
    ...        &     ...        &  ...  &   ...  & ...& ... &     ...  &  ...   &  ...  &   ...  &  ... &  ...  &  \\
\hline
\end{tabular}
\end{center}
\end{table*}

\begin{table*}
\begin{center}
\caption[]{Spatial distribution of the Cepheids. The table lists the Galactic coordinates, the heliocentric distance, the pulsation period and pulsation mode, the azimuthal angle $\phi$, and the Galactocentric distance \Rg{}. The metallicity [Fe/H] is recalled in the last column.
(Only the first ten lines of the table are shown. The full table is available in electronic form)}
\label{abund1}
\begin{tabular}{rrrrrrrr}
\hline
        Star  &     l   &     b   & Distance & Period, pulsation mode & $\phi$ &   Rg   & [Fe/H]  \\
              & (deg) & (deg) &   (pc)  &     (day)       & (deg) &  (kpc) &  dex   \\
\hline
        RW CMa&  232.04 &  --3.81 &    3051 &   5.7297117 F   & --13.47   & 10.31  &   0.05 \\
      V384 CMa&  230.28 &  --5.38 &    4679 &   4.2059423 F   & --17.85   & 11.69  &   0.02 \\
        CC Car&  289.37 &  --1.59 &    4515 &   4.7598281 F   & --32.62   &  7.90  &   0.18 \\
        CR Car&  285.66 &  --0.37 &    4740 &   9.7588957 F   & --33.60   &  8.25  &   0.14 \\
        CT Car&  287.63 &  --2.77 &   10083 &  18.0608858 F   & --62.02   & 10.87  &   0.07 \\
        FN Car&  289.60 &  --0.12 &    3811 &   4.5856135 F   & --27.59   &  7.75  &   0.13 \\
        FQ Car&  290.91 &  --0.35 &    4261 &  10.2739900 F   & --30.98   &  7.73  &   0.17 \\
        FZ Car&  288.39 &    0.31 &    3962 &   3.5780713 1O  & --28.58   &  7.86  &   0.13 \\
        GI Car&  290.26 &    2.54 &    1912 &   4.4307267 1O  & --13.46   &  7.70  &   0.03 \\
      V690 Car&  280.59 &  --3.31 &    3368 &   4.1505759 F   & --23.69   &  8.22  &   0.20 \\
           ...&  ...    &   ...   &     ... &     ...         &   ...     &  ...   &    ...  \\
\hline
\end{tabular}
\end{center}
\end{table*}

\begin{table*}
\begin{center}
\caption[]{Abundances of Cepheids (C -- Mn). (Only the first ten lines of the table are shown. The full table is available in electronic form)}
\label{abund2}
\begin{tabular}{rrrrrrrrrrrrrrr}
\hline
              &   C  &    N &    O &    Na&    Mg&    Al&    Si&    S &    Ca&    Sc&   Ti &   V  &   Cr &   Mn   \\
\hline
       RW CMa &--0.46&  0.46&--0.04&  0.31&  0.02&  0.19&  0.11&  0.09&  0.03&  0.09&  0.09&--0.12&--0.04&--0.17  \\
     V384 CMa &--0.51&  0.46&--0.28&  0.12&   -- &  0.06&--0.02&--0.02&--0.05&  0.07&--0.01&--0.22&--0.10&--0.29  \\
       CC Car &--0.17&  0.47&--0.11&  0.30&  0.02&  0.33&  0.22&  0.36&  0.14&   -- &  0.08&--0.10&  0.00&--0.07  \\
       CR Car &--0.31&  0.46&  0.12&  0.26&--0.14&  0.19&  0.19&  0.13&  0.15&  0.22&  0.18&  0.03&--0.01&--0.11  \\
       CT Car &--0.32&  0.40&--0.01&  0.25&  0.10&  0.15&  0.15&  0.13&--0.07&   -- &--0.07&--0.09&--0.10&--0.11  \\
       FN Car &--0.33&  0.37&  0.09&  0.31&  0.07&  0.22&  0.20&  0.18&  0.15&  0.07&  0.13&--0.08&--0.03&--0.08  \\
       FQ Car &--0.18&  0.53&  0.15&  0.39&  0.07&  0.27&  0.26&  0.32&  0.08&   -- &  0.09&--0.03&  0.03&--0.03  \\
       FZ Car &--0.21&  0.51&--0.11&  0.29&  0.09&  0.21&  0.18&  0.23&  0.14&  0.17&  0.15&--0.13&  0.00&--0.09  \\
       GI Car &--0.33&  0.34&--0.14&  0.18&--0.01&  0.13&  0.12&  0.30&  0.06&  0.28&  0.10&--0.02&--0.09&--0.21  \\
     V690 Car &--0.18&  0.46&  0.13&  0.38&  0.12&  0.32&  0.23&  0.26&  0.11&  0.23&  0.10&--0.07&  0.01&--0.09  \\
          ... &  ... &  ... &  ... &  ... &  ... &  ... &  ... &  ... &  ... &  ... &  ... &  ... &  ... &  ...   \\
\hline
\end{tabular}
\end{center}
\end{table*}

\begin{table*}
\begin{center}
\caption[]{Abundances of Cepheids (Fe -- Gd). (Only the first ten lines of the table are shown. The full table is available in electronic form)}
\label{abund3}
\begin{tabular}{rrrrrrrrrrrrr}
\hline
    Star     &   Fe &    Co&    Ni&   Y  &   Zr &   La &   Ce &   Pr &   Nd &    Sm&   Eu &    Gd\\
\hline
       RW CMa&  0.05&  0.10&--0.09&  0.07&--0.09&  0.03&  0.05&--0.15&--0.03&--0.13&  0.03&   -- \\
     V384 CMa&--0.01&--0.14&--0.16&  0.08&--0.07&  0.05&  0.07&--0.19&  0.12&  0.00&  0.03&   -- \\
       CC Car&  0.18&--0.11&  0.00&  0.22&  0.17&  0.10&  0.02&--0.18&  0.02&  0.00&  0.18&   -- \\
       CR Car&  0.14&  0.08&  0.01&  0.17&  0.22&  0.13&  0.09&--0.06&  0.07&--0.02&  0.16&  0.15\\
       CT Car&  0.07&--0.14&--0.09&  0.20&--0.08&  0.28&  0.00&  0.01&  0.09&--0.02&  0.11&--0.15\\
       FN Car&  0.13&--0.08&--0.01&  0.16&  0.17&  0.08&  0.05&--0.13&  0.06&--0.06&  0.11&   -- \\
       FQ Car&  0.17&--0.02&  0.04&  0.20&  0.03&  0.22&  0.10&--0.05&  0.15&  0.13&  0.25&   -- \\
       FZ Car&  0.13&  0.02&--0.01&  0.28&--0.05&  0.14&  0.01&--0.16&  0.07&  0.05&  0.16&   -- \\
       GI Car&  0.03&--0.06&--0.12&  0.25&  0.14&  0.14&  0.08&--0.20&  0.06&  0.02&  0.16&   -- \\
     V690 Car&  0.20&--0.08&--0.02&  0.28&--0.04&  0.21&  0.11&--0.05&  0.15&  0.05&  0.27&   -- \\
        ...  &   ...& ...  &  ... &  ... &  ... &  ... &  ... &  ... &  ... &  ... &  ... &   ...\\
\hline
\end{tabular}
\end{center}
\end{table*}

\par The abundances of different elements were derived in the LTE approximation
using atmosphere models interpolated for the specific atmospheric parameters of each individual star within the grid of models by \cite{ck04}. We discarded strong lines (with EWs$>$150 m\AA) due to noticeable damping effects. The list of the lines measured is given
in \cite{Lemasle2015}. The oscillator strengths, log gf, are adopted from the Vienna Atomic Line Database (VALD) \citep[][version 2018]{Kupka1999}. The solar abundances are taken from \cite{Asplund2009}.  

\par To estimate the error budget on the abundances derived, we proceeded as follows:
\begin{itemize}[nosep,labelsep=0.5em]
    \item[-] We derived \Teff\ typically from 50 to 70 line-depth ratios, which resulted in standard deviations of the mean of individual temperatures of $\Delta$\Teff = $\pm100~K$, which we adopted as our uncertainty on \Teff.
    \item[-] We enforced the ionisation balance for iron by limiting at 0.05~dex the spread between the total iron abundance derived from the Fe~{\sc i} and Fe~{\sc ii} lines. This corresponds to an uncertainty of $\Delta$\logg = $\pm0.2$ dex, which we adopted in our error budget.  
    \item[-] A variation $\Delta$\Vt = $\pm0.3$ \kms{} results in a significant slope of the relationship between [Fe/H] derived from the Fe~{\sc i} lines and their equivalent widths. Since we imposed a flat relation (no slope), we adopted the above value as the uncertainty on the microturbulent velocity. 
\end{itemize}
\par We computed the abundances with deliberately over- or under-estimated values for a given atmospheric parameter and computed the total uncertainty as the sum in quadrature of the uncertainties relative to a single parameter. Such a procedure overestimates the uncertainties in the final abundances due to uncertainties in the atmospheric parameters, as the latter are correlated. We repeated the exercise for two representative Cepheids and the uncertainties on the abundances are reported in Table~\ref{unc_param}).

\tiny
\begin{table*}
\centering
\caption{Uncertainties in the final abundances due to uncertainties in the atmospheric parameters. Cols. 2, 3, 4 indicate respectively how the abundances are modified (mean values) when they are computed with over- or underestimated values of \Teff ($\pm$100 K), \logg ($\pm$0.2 dex), or \Vt ($\pm$0.3 dex). The sum in quadrature of the differences is adopted as the uncertainty in the abundances due to the uncertainties in the atmosphere parameters}
\begin{tabular}{lrrrrcrrrr}
\hline
\multicolumn{1}{c}{}&\multicolumn{4}{c}{Error budget for BB Sgr, P=6.63308 d}&\multicolumn{1}{c}{}&\multicolumn{4}{c}{Error budget for V5567 Sgr, P=9.76316}\\ \cmidrule{2-5}\cmidrule{7-10}
\multicolumn{1}{c}{Ion}     & \multicolumn{1}{c}{$\Delta$\Teff} & \multicolumn{1}{c}{$\Delta$\logg} &  \multicolumn{1}{c}{$\Delta$\Vt}  & \multicolumn{1}{c}{Total} & & \multicolumn{1}{c}{$\Delta$\Teff} & \multicolumn{1}{c}{$\Delta$\logg} &  \multicolumn{1}{c}{$\Delta$\Vt}  & \multicolumn{1}{c}{Total} \\
        & {\small ($\pm$ 100 K)}   & {\small  ($\pm$ 0.2 dex)} & {\small($\pm$ 0.3 km~s$^{-1}$)}  &     \multicolumn{1}{c}{dex}  & & {\small ($\pm$ 100 K)}   & {\small  ($\pm$ 0.2 dex)} & {\small($\pm$ 0.3 km~s$^{-1}$)}  &     \multicolumn{1}{c}{dex}     \\\cmidrule{2-5}\cmidrule{7-10}
C   {\sc i}   &  --0.07&    0.04&  --0.01&  0.08& &  --0.05&    0.09&  --0.02&  0.10\\
N   {\sc i}   &  --0.11&    0.03&  --0.02&  0.12& &  --0.09&    0.12&  --0.02&  0.15\\
O   {\sc i}   &    0.03&    0.08&  --0.01&  0.09& &    0.00&    0.08&  --0.01&  0.08\\
Na  {\sc i}   &    0.06&  --0.01&  --0.03&  0.07& &    0.05&  --0.01&  --0.03&  0.06\\
Mg  {\sc i}   &  --0.07&    0.01&    0.08&  0.11& &    0.04&    0.00&  --0.04&  0.06\\
Al  {\sc i}   &    0.05&  --0.01&  --0.03&  0.06& &    0.04&    0.00&  --0.02&  0.04\\
Si  {\sc i }  &    0.05&    0.00&  --0.02&  0.05& &    0.06&    0.02&  --0.03&  0.07\\
S   {\sc i }  &  --0.06&    0.04&  --0.02&  0.07& &  --0.03&    0.08&  --0.02&  0.09\\
Ca  {\sc i }  &    0.07&  --0.01&  --0.03&  0.08& &    0.06&  --0.01&  --0.06&  0.09\\
Sc  {\sc ii}  &    0.01&    0.08&  --0.04&  0.09& &    0.03&    0.09&  --0.05&  0.11\\
Ti  {\sc i }  &    0.11&    0.00&  --0.02&  0.11& &    0.09&  --0.02&  --0.01&  0.09\\
Ti  {\sc ii}  &    0.00&    0.08&  --0.04&  0.09& &    0.01&    0.08&  --0.04&  0.09\\
V   {\sc i }  &    0.12&    0.01&  --0.01&  0.12& &    0.10&  --0.02&  --0.01&  0.10\\
V   {\sc ii}  &    0.00&    0.08&  --0.01&  0.08& &    0.02&    0.08&  --0.01&  0.08\\
Cr  {\sc i }  &    0.07&    0.00&  --0.02&  0.07& &    0.06&  --0.01&  --0.02&  0.06\\
Cr  {\sc ii}  &  --0.03&    0.07&  --0.05&  0.09& &    0.00&    0.10&  --0.06&  0.12\\
Mn  {\sc i }  &    0.07&    0.00&  --0.03&  0.08& &    0.07&  --0.01&  --0.03&  0.08\\
Fe  {\sc i }  &    0.08&    0.00&  --0.02&  0.08& &    0.07&  --0.01&  --0.03&  0.08\\
Fe  {\sc ii}  &  --0.03&    0.07&  --0.02&  0.08& &    0.00&    0.10&  --0.03&  0.10\\
Co  {\sc i }  &    0.11&    0.01&  --0.01&  0.11& &    0.08&  --0.02&  --0.01&  0.08\\
Ni  {\sc i }  &    0.09&    0.01&  --0.02&  0.09& &    0.08&  --0.01&  --0.03&  0.09\\
Y   {\sc ii}  &    0.00&    0.07&  --0.04&  0.08& &    0.03&    0.08&  --0.04&  0.09\\
Zr  {\sc ii}  &    0.00&    0.07&  --0.04&  0.08& &    0.03&    0.10&  --0.04&  0.11\\
La  {\sc ii}  &    0.03&    0.08&  --0.03&  0.09& &    0.05&    0.08&  --0.02&  0.10\\
Ce  {\sc ii}  &    0.02&    0.07&  --0.01&  0.07& &    0.03&    0.07&  --0.03&  0.08\\
Pr  {\sc ii}  &    0.03&    0.08&  --0.01&  0.09& &    0.05&    0.07&  --0.01&  0.09\\
Nd  {\sc ii}  &    0.02&    0.07&  --0.02&  0.08& &    0.05&    0.08&  --0.03&  0.10\\
Sm  {\sc ii}  &    0.03&    0.08&  --0.01&  0.09& &    0.05&    0.08&  --0.01&  0.09\\
Eu  {\sc ii}  &    0.01&    0.07&  --0.02&  0.07& &    0.03&    0.08&  --0.02&  0.09\\
Gd  {\sc ii}  &    0.02&    0.08&    0.00&  0.08& &    0.03&    0.07&  --0.01&  0.08\\
\hline
\end{tabular}
\label{unc_param}
\end{table*}

\normalsize

\begin{figure} 
\resizebox{\hsize}{!}{\includegraphics{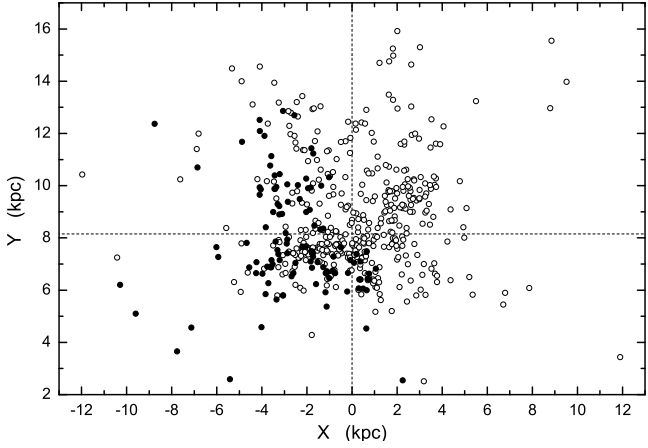}}
\caption{
Locations of target Cepheids in the Galactic plane as viewed
from the North Galactic Pole. The position of the Sun is at
the intersection of the dashed lines. The Galactic center is at (X,Y) = (0, 0). Open and filled circles
are used to mark the Cepheids adopted from Luck (2018) and
our new SALT observations, respectively.}
\label{map}
\end{figure}

\begin{figure} 
\resizebox{\hsize}{!}{\includegraphics{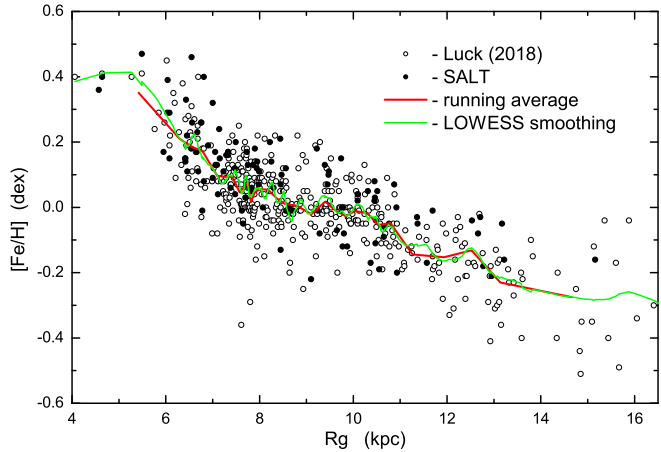}}    
\caption{The radial distribution of the iron abundance. The filled circles indicate the Cepheids reported in this paper. The Cepheid variables from \citet{Luck2018} are marked
 with open circles, their Galactocentric distances were recomputed for \Rsun = 8.15 kpc. The solid red line represents a running average while the green line results
from a LOWESS smoothing (the degree of polynomial = 2; bandwidth = 0.07).}
\label{Fe-Rg1}
\end{figure}

\begin{figure} 
\resizebox{\hsize}{!}{\includegraphics{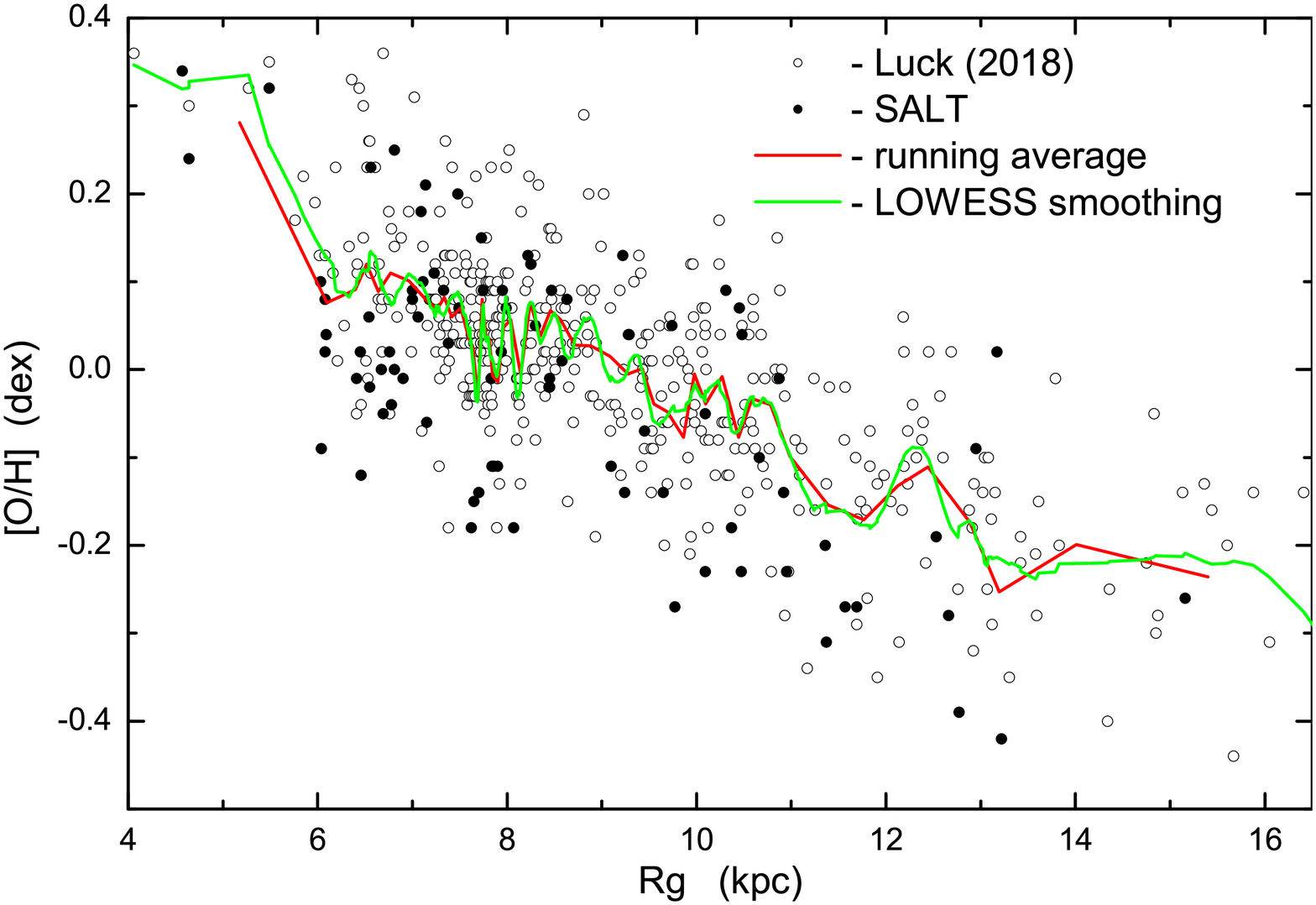}}    
 \caption{The same as in Fig \ref{Fe-Rg1} but for [O/H].}
\label{Oxygen}
\end{figure}

\begin{figure} 
\resizebox{\hsize}{!}{\includegraphics{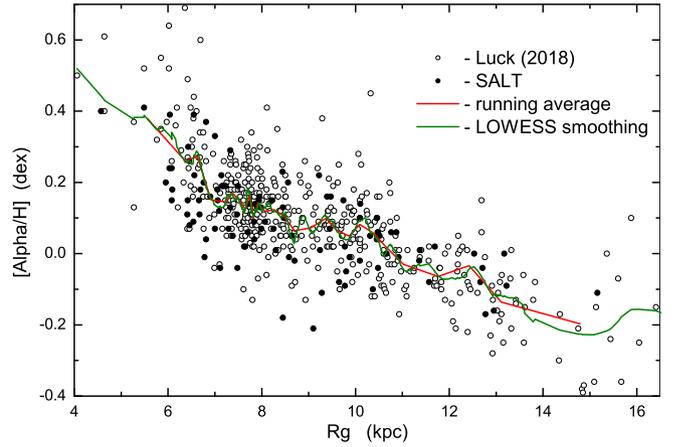}}    
 \caption{The same as in Fig \ref{Fe-Rg1} but for $\alpha$-elements
(that is, the average abundances of Si + Mg + Ca) in Cepheids,
 normalised to the solar value.}
\label{Alpha}
\end{figure}

\begin{figure} 
\resizebox{\hsize}{!}{\includegraphics{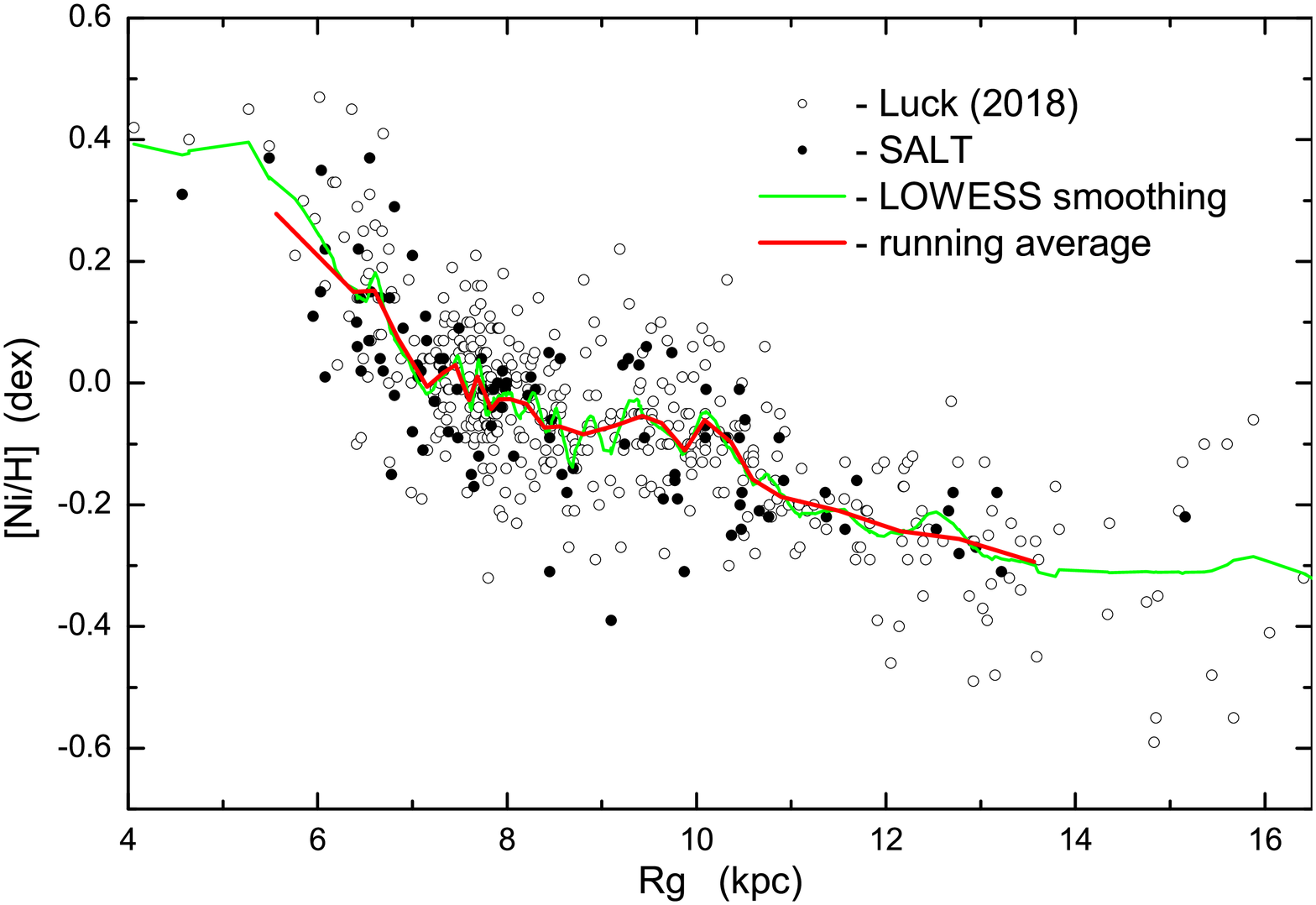}}    
\caption{The same as in Fig \ref{Fe-Rg1} but for [Ni/H].
}
\label{Ni}
\end{figure}

\begin{figure} 
\resizebox{\hsize}{!}{\includegraphics{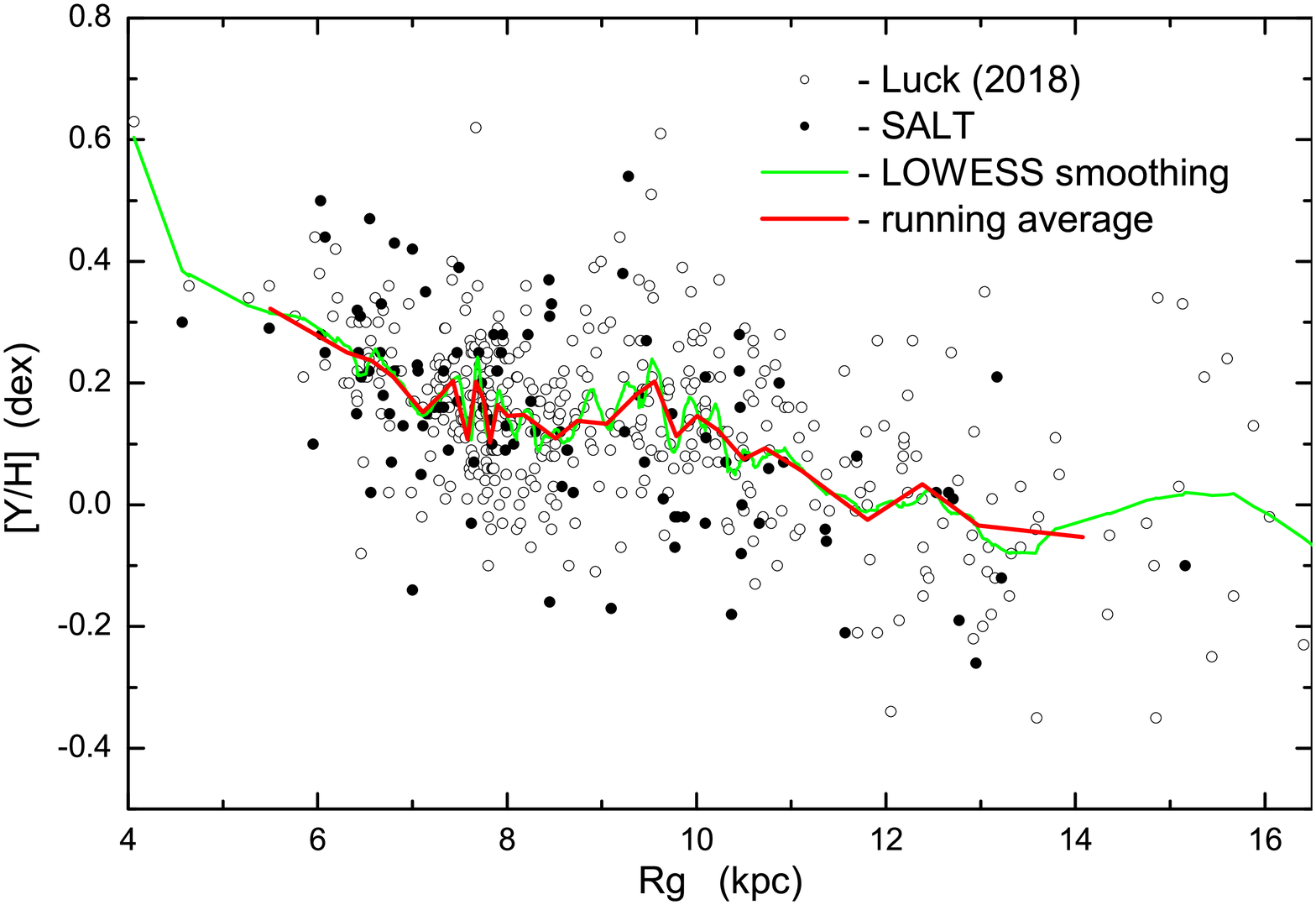}}
\caption{The same as in Fig \ref{Fe-Rg1} but for [Y/H].
}
\label{Y}
\end{figure}

\begin{figure} 
\resizebox{\hsize}{!}{\includegraphics{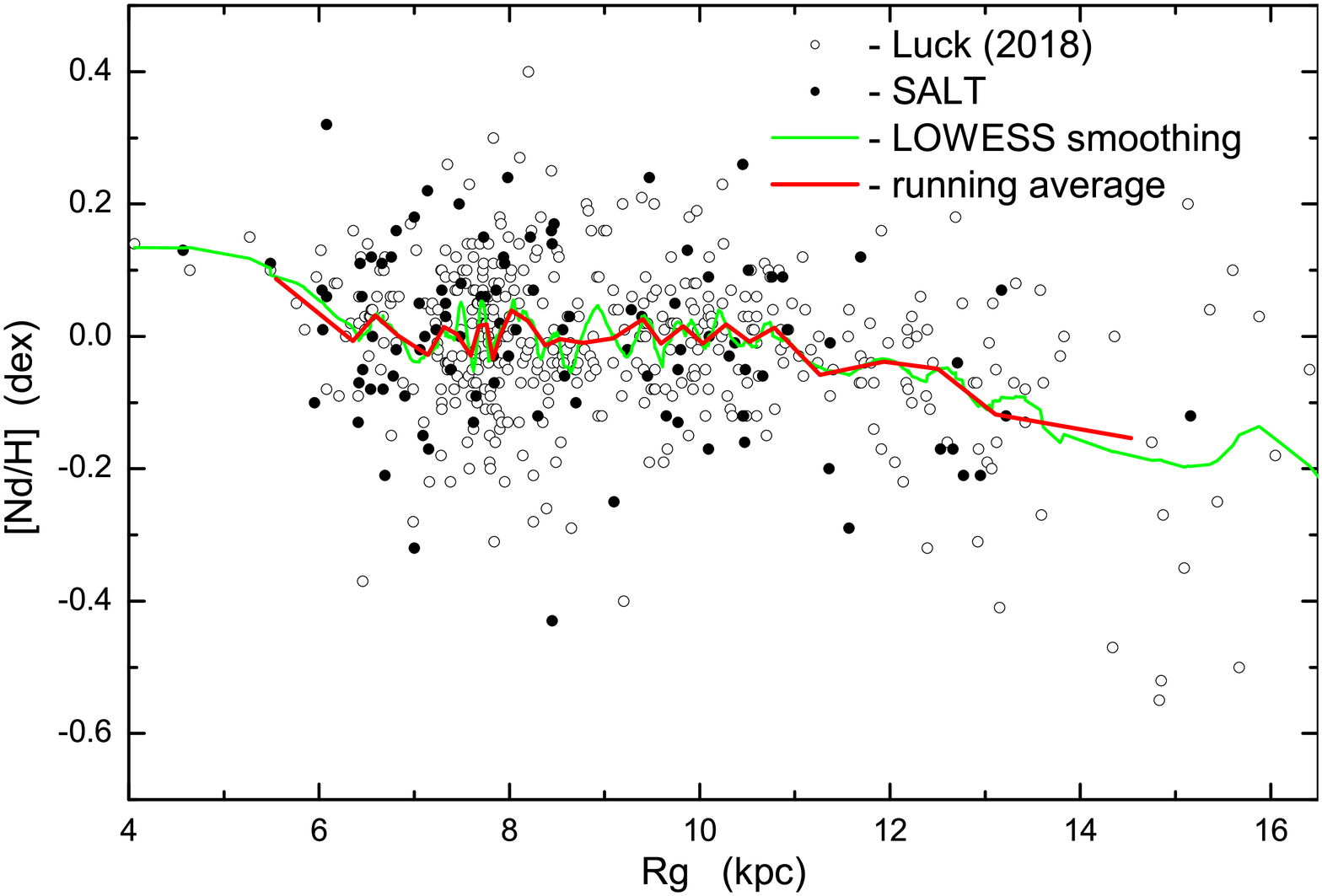}}
\caption{The same as in Fig \ref{Fe-Rg1} but for [Nd/H].
}
\label{Nd}
\end{figure}

\section{The Galactic radial abundance gradients}
\label{results}

\par In this section, we study the radial abundance gradient for different groups of elements. As is customary for gradient studies, we display our measurements in the [Fe/H] vs Galactocentric distance (\Rg) plane. The possible azimuthal dependence of the gradients are  discussed in the next section.
We complement our new chemical abundances (105 Cepheids) by literature values compiled by \cite{Luck2018}, who conducted a similar spectroscopic analysis.

\subsection{Distances}

\par For our new sample as well as for the Cepheids from the literature, we have computed the Galactocentric distances by adopting a distance to the Galactic centre of \Rsun = 8.15 kpc (\citealt{Gravity2019, Reid2019}) and heliocentric distances to Cepheids from \cite{Skowron2019}. \cite{Skowron2019} used mid-infrared Spitzer \citep[][]{Churchwell2009} and WISE \citep[][]{Mainzer2011} photometry together with the mid-infrared period-luminosity relations derived by \cite{Wang2018} and the extinction maps of \cite{Bovy2016}.

\par Our sample includes $\sim$10 Cepheids with \Rg\,$\lesssim$ 6\,kpc (25 with \Rg\,$<$ 7\.kpc), improving the sampling of the inner Galactic region. It also includes 14 Cepheids with \Rg\,$>$ 11\,kpc. Such stars are important as the outer disc remains poorly sampled compared to the Solar neighborhood. Moreover, these new distant Cepheids are mostly located in the fourth Galactic quadrant (see Fig.~\ref{map}), which has been only partially explored from the Cepheids point of view. The relevant information regarding the spatial location of the Cepheids is gathered in Table~\ref{abund1}. 

\subsection{Radial abundance gradients}
\label{rad_grad}

\par The radial iron gradient is shown in Fig.~\ref{Fe-Rg1}. Our new measurements overlap perfectly previous results in the [7-10]~kpc range. They confirm a steeper slope in the inner disc as already proposed by \citet{Andrievsky2002b,Pedicelli2010}. [Fe/H] caps at $\approx$ +0.4--+0.5~dex, and there are hints that there is a plateau around those values towards the inner disc, as already suggested by \citet[][]{Martin2015,Inno2019}. The few Cepheids known to be located in the vicinity of the Galactic center have been spectroscopically analyzed by \citet[][]{Kovtyukh2019} 
\par Since oxygen is predominantly released in type II supernovae (SNe~II), it is important to trace the radial abundance gradient of oxygen in order to investigate the impact of the spiral arms on the Galactic chemical enrichment. Indeed, the progenitors of SNe~II are young massive stars, and most of them explode before having left the spiral arm in which they are born \citep{Acharova2005}. However, oxygen is difficult to measure in Cepheids because the few oxygen lines available in Cepheids' atmospheres are blended or strongly affected by NLTE effects \citep[][]{Korotin2014,Vasilyev2019}. The oxygen radial gradient, displayed in Fig.~\ref{Oxygen}, nevertheless follows the iron gradient, albeit with a much larger scatter.
\par The same holds for the $\alpha$-elements gradients, for which we show the radial distribution of the average abundances of the elements Si, Mg and Ca (with equal weights) in Fig.~\ref{Alpha}. As expected, the oxygen and $\alpha$-element radial abundance gradients, are within the 
errors quite similar. Similarly to oxygen, sulphur abundances show a large scatter at a given \Rg{} due to the small number (and weakness) of S lines available in the spectral range covered by our spectra.
\par It is no surprise that iron--peak elements follow the radial distribution of [Fe/H], as exemplified by the [Ni/H] gradient shown in Fig.~\ref{Ni}
\par The large scatter in the distribution of neutron-capture elements as a function of \Rg\ prevents us from drawing a stronger conclusion than an overall decrease from the inner to the outer disc, as already found by e.g., \citet[][]{Lemasle2013,daSilva2016}. We note that the only two europium lines in the optical spectra of Cepheids are usually weak, and they are affected by hyperfine structure splitting (hfs). However hfs corrections were found to be negligible by \cite{daSilva2016} and therefore cannot account for the observed scatter. The same authors also reported negligible corrections for the three Y lines for which hfs data were available from \citet[][]{McWilliam2013}. The [Y/H] radial gradient is shown in Fig.~\ref{Y}. Similarly, the laboratory transition probabilities provided by \citet[][]{DenHartog2003} indicate no evident HFS structure for Nd {\sc ii}. Fig.~\ref{Nd} displays the [Nd/H] radial gradient.
\par Abundances from C to Mn are listed in Table~\ref{abund2} and abundances from Fe to Gd are listed in Table~\ref{abund3}.

\begin{figure*}
\begin{tabular}{cc}
\includegraphics[width=8cm]{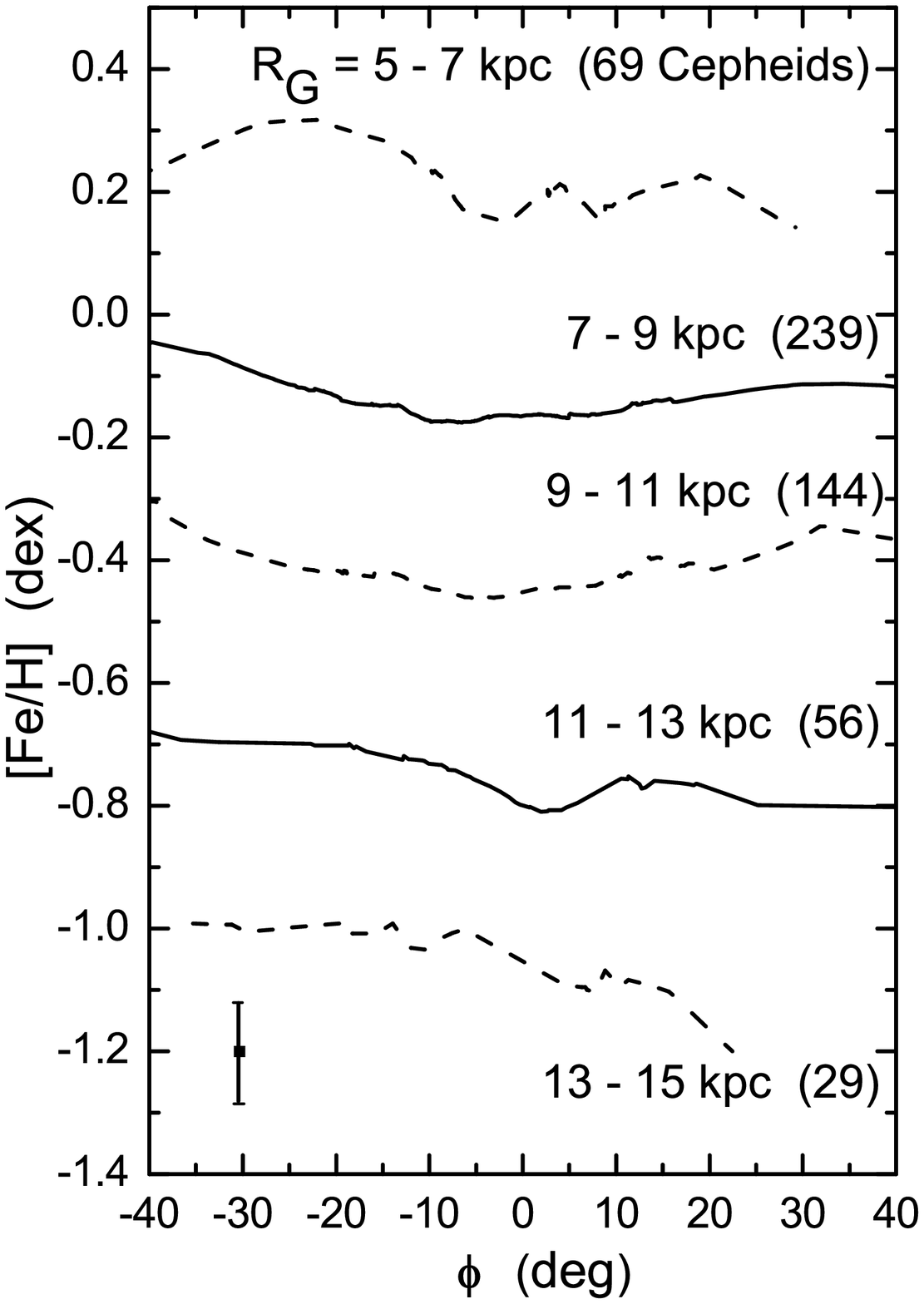}\includegraphics[width=8cm]{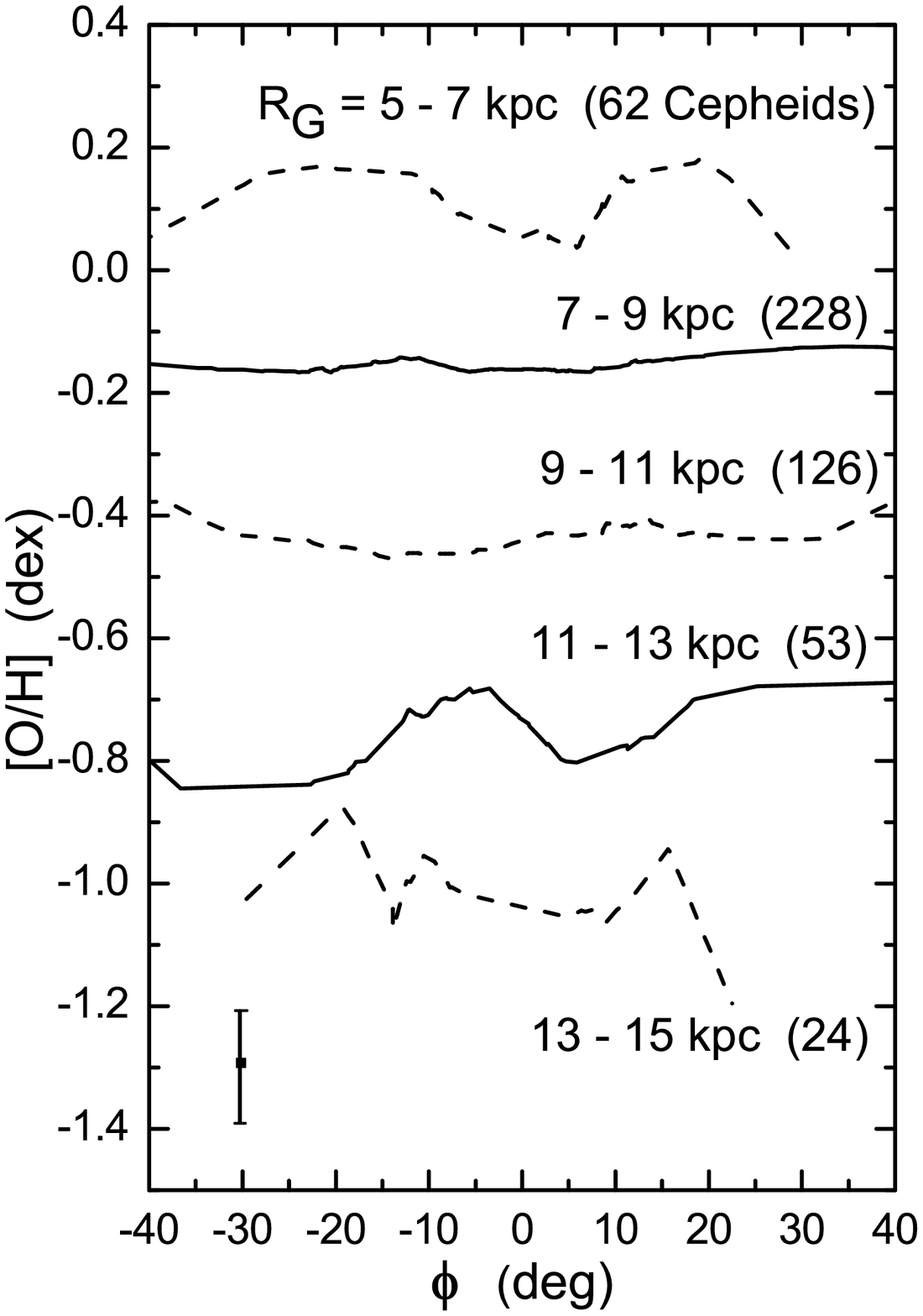}\\
\end{tabular}
\caption{
LOWESS smoothing of [Fe/H] and [O/H] for the azimuthal coordinate $\phi$
at different \Rg.
The [El/H] ratios have been
arbitrarily shifted for the sake of clarity. The total number of Cepheids in each annulus is indicated. Representative uncertainties on the {\it individual} Cepheids abundances are shown at the bottom left of each panel.}
\label{5alpha}
\end{figure*}

\subsection{The azimuthal abundance gradient}

\par Figure \ref{5alpha} illustrates the azimuthal variations in the abundance ratios at different Galactocentric distances. The azimuth, $\phi$, is defined as the angle between the Galactocentric radius containing a given Cepheid and the reference radius ($\phi$=0) containing both the Sun and the Galactic centre. $\phi$ increases with the Galactocentric longitude. The azimuthal variations have been smoothed using a
LOWESS algorithmn \citep[Locally weighted scatterplot
smoothing,][]{Cleveland1979}
using a first degree polynomial and a bandwidth of 0.30.
This means that 30\% of the sample is used for each local fit, with
the weight of any considered data point strongly decreasing with its
distance to the point on the curve being fitted. The total number of
Cepheids in each annulus, ranging from a few tens to a few hundreds,
is indicated in the figure. Given the difficulties to measure oxygen
(see Sect.~\ref{rad_grad}), the numbers are slightly larger for [Fe/H]
than for [O/H].
The azimuthal variations of [Fe/H] remain within 0.2~dex at all radii.
They are larger for the inner (5--7~kpc) and outer (13--15~kpc) annuli,
and minimal ($\approx$0.1~dex) for the 9--11 and 11--13~kpc annuli.
A similar pattern can be observed for the azimuthal variations of
[O/H], with large fluctuations ($\approx$0.2~dex) in the inner and
outer Galactic disc, and minimal variations ($\approx$0.05~dex)
within 7--9 and 9--11~kpc.

Since our azimuthal variations are measured over a single annulus
encompassing a fixed range of Galactocentric distances, a given
annulus might overlap more than one spiral arm. We speculate that
the azimuthal inhomogeneities could occur in regions where an annulus
transitions from one spiral arm to another, for instance from the Norma
to the Scutum-Centaurus arm in the inner disk, or from the Local to
the Sagittarius-Carina arm, when comparing to the model by \citet[][]{Hou2021}.
Under this hypothesis, the limited variations in the 9-11\,kpc ring could be
explained by the fact that the corresponding annulus follows the Perseus arm.
If this hypothesis turns out to be correct, it could imply that the chemical
composition of Cepheids remains similar within a given spiral arm, but
varies when compared to Cepheids from another arm. Unfortunately, it
is for now impossible to attribute with certainty a given Cepheid to
a specific spiral arm.

\par \citet[][]{Davies2009,Origlia2013} already reported large scale azimuthal variations in the inner galaxy using red supergiants (RSG) in the Scutum clusters, located at the end of the bar. \citet[][]{Genovali2013} noted that although the age difference between both tracers is minimal (a few Myr to a
few tens of Myrs), Cepheids attain supersolar iron abundances in the inner disc (0.4--0.5 dex), while the Scutum RSGs have subsolar metallicities (-0.20-- -0.3 dex), hence a difference larger than 0.5 dex. These variations are usually attributed to the Galactic bar, they were later supported by \citet[][]{Inno2019}.

\par A number of recent studies using integral field spectrographs updated the study of azimuthal variations in the properties of nearby galaxies, which was previously achieved via long-slit spectroscopy: \cite{Li2013} detected azimuthal variations in the oxygen abundance in the spiral galaxy M~101 (NGC~5457). Both \citet[][]{Sanchez2015} and \citet[][]{Sanchez-Menguiano2016} reported azimuthal variations in [O/H] in NGC~6754. They found larger inhomogeneities in the outer regions of this galaxy. \citet[][]{Ho2017} found azimuthal variations (with respect to the oxygen radial gradient) of the order of $\approx$0.2~dex associated with the two spiral arms of NGC~1365 and \citet[][]{Ho2018} reported similar variations of 0.06 dex in NGC~2997. 
\par It is expected that the spiral arm structure plays an important role in homogenising the interstellar medium (ISM) \citep[e.g.,][]{Kreckel2019}. \citet[][]{Ho2017} suggested that chemical enrichment takes place in tiny gas pockets in the interarm region, and that the ISM gets homogenised when crossing the spiral arm. In their analysis of the galaxy HCG~91c, \citet{Vogt2017} found both localised variations associated with individual H~{\sc ii} regions, and extended structures at the boundaries of the spiral arms. They concluded that the enrichment of the ISM preferentially takes place along the spiral structure rather than in the interarm regions. 
\par It should be noted that galaxies with large azimuthal variations represent a very small sample with respect to the overall population \citep[e.g.,][]{Sanchez-Menguiano2018}, and the inhomogeneities detected are usually of small amplitude. For instance \citet[][]{Zinchenko2016} reported typical asymmetries $<$\,0.05 dex in 88 galaxies of the Calar Alto Legacy Integral Field Area survey (CALIFA) Data Release 2. This raises the question of the origin of the larger ones and might indicate that they are related to localised events rather than to the secular evolution of the galaxy considered.
\par \cite{Spitoni2019} presented a 2D chemical evolution model of the Milky Way that is capable of tracing azimuthal variations of chemical abundances in the Galactic disc. Density fluctuations produce
significant azimuthal variations in [O/H], of the order of
0.1 dex around the mean. Such variations are more obvious in the outer disc.
Our results support qualitatively the findings of \cite{Spitoni2019}. Similar conclusions were reached by \citet{Spitoni2019} when comparing the outcome of their models to the Cepheids data of \citet{Genovali2014}. Our results are also in agreement with the model predictions by \citet[][]{Molla2019}.
However, the variations in [O/H] are slightly larger in our observational data than in the
\citet{Spitoni2019}, and especially, the \citet{Molla2019} model, which we
attribute to the (significant) uncertainties in the [O/H] abundances of
Cepheids. Using a simple analytical spiral arm model, \citet[][]{Molla2019}
mention that the azimuthal variations in their models are smaller than the
typical uncertainties associated with oxygen abundance tracers, and that
they reach detectable values ($\approx$0.1~dex, (peak to peak) only in the outer regions of the disc. \cite{Spitoni2019} specify that the largest fluctuations in azimuthal abundance gradients occur near the corotation radius of the spiral pattern, where chemical enrichment becomes more efficient because of the absence of relative motion between the gas and the spiral arm. \citet[][]{Solar2020} analyzed oxygen gradients in 106 resolved spiral galaxies selected from the high resolution Evolution and Assembly of GaLaxies and
 their Environments (EAGLE) simulations \citep[][]{Schaye2015}. In this sample, they found a large scatter for the [O/H] gradient measured at a random azimuthal direction when compared to its global average over the disc, which they interpret as an evidence significant azimuthal variations. 

\section{Conclusion}

\par We report the chemical composition (25 elements) of 105 classical Cepheids. Abundances have been derived from 122 high-resolution SALT spectra observed in the context of the MAGIC project. The Galactocentric distances of the Cepheids are taken from the literature, they are based on mid-infrared photometry. Our new sample contains 25 Cepheids with \Rg\,$<$7\,kpc (14 with \Rg\,$<$\,6.5\,kpc), and 14 Cepheids with \Rg\,$>$\,11\,kpc, thus significantly improving the sampling in the inner and outer disc. Our results are in line with Cepheids literature studies concerning radial abundance gradients of iron, $\alpha$, or neutron-capture elements. We focus on the azimuthal variations of the oxygen abundance [O/H]. Since such variations are found to be negligible (typically $<$\,0.05~dex) in external spirals, the Milky Way seems to be one of the few galaxies with noticeable [O/H] azimuthal inhomogeneities. We find that they are the largest in the inner Galaxy, where they are usually attrib

uted to perturbations induced by the rotating bar, and in the outer disc. In these outer regions, our results support similar results obtained in a few nearby spiral galaxies as well as the outcome of chemo-dynamical evolution models by e.g., \citet[][]{Spitoni2019,Molla2019}.

\section{Acknowledgements}
The authors thank the anonymous referee for her/his careful reading of the paper and for her/his comments, which improved the clarity of our manuscript.
This work was funded by the Deutsche Forschungsgemeinschaft
(DFG, German Research Foundation) -- Project-ID 138713538 -- SFB 881
(``The Milky Way System'', subproject A05). A.K. acknowledges the National
Research Foundation of South Africa.

\section{Data Availability}

The SALT data are available in the SALT
Archive at https://astronomers.salt.ac.za
after the proprietary period.


\begin{thebibliography}{99}
\bibitem[\protect\citeauthoryear{Acharova et al.}{2005}]{Acharova2005}
Acharova I.~A., L{\'e}pine J.~R.~D., Mishurov Yu.~N.,
2005, MNRAS, 359, 819
\bibitem[\protect\citeauthoryear{{Andrievsky}, {Kovtyukh}, {Luck},
{L{\'e}pine}, {Bersier}, {Maciel}, {Barbuy}, {Klochkova}, {Panchuk} \&
{Karpischek}}{{Andrievsky} et~al.}{2002a}]{Andrievsky2002a}
{Andrievsky} S.~M.,  {Kovtyukh} V.~V.,  {Luck} R.~E.,  {L{\'e}pine} J.~R.~D.,
{Bersier} D.,  {Maciel} W.~J.,  {Barbuy} B.,  {Klochkova} V.~G.,  {Panchuk}
V.~E.,    {Karpischek} R.~U.,  2002a, \aap, 381, 32   
\bibitem[\protect\citeauthoryear{{Andrievsky}, {Bersier}, {Kovtyukh}, {Luck}, {Maciel}, {L{\'e}pine} \& {Beletsky}}{{Andrievsky} et~al.}{2002b}]{Andrievsky2002b}
{Andrievsky} S.~M.,  {Bersier} D.,  {Kovtyukh} V.~V.,  {Luck} R.~E.,  {Maciel} W.~J.,  {L{\'e}pine} J.~R.~D., {Beletsky} Y.~V.,  2002b, \aap, 384, 140
\bibitem[{{Asplund} {et~al}\mbox{.}(2009){Asplund}, {Grevesse}, {Sauval}, \&
  {Scott}}]{Asplund2009}
{Asplund} M., {Grevesse} N., {Sauval} A.~J., {Scott} P., 2009, \araa, 47, 481
\bibitem[\protect\citeauthoryear{Balser et al}{2011}]{Balser2011}
Balser D.~S., Rood R.~T., Bania T.~M., Anderson L.~D.,
2011, ApJ, 738, 27
\bibitem[\protect\citeauthoryear{Balser et al}{2015}]{Balser2015}
Balser, D.~S., Wenger, T.~V., Anderson, L.~D.,  Bania, T.~M.,
2015, ApJ, 806, 199
\bibitem[\protect\citeauthoryear{Bovy et al.}{2016}]{Bovy2016} Bovy J., Rix H.-W., Green G.~M., Schlafly E.~F., Finkbeiner D.~P., 2016, ApJ, 818, 130. doi:10.3847/0004-637X/818/2/130
\bibitem[\protect\citeauthoryear{Buckley, Swart \& Meiring}{2006}]{Buck2006}
Buckley D.~A.~H., Swart G.~P., Meiring J.~G., 2006, SPIE, 6267
\bibitem[\protect\citeauthoryear{Castelli \& Kurucz}{2004}]{ck04}
Castelli F., Kurucz R.L., 2004, arXiv:astro-ph/0405087
\bibitem[\protect\citeauthoryear{Cescutti et al.}{2007}]{Cescutti2007} Cescutti G., Matteucci F., Fran{\c{c}}ois P., Chiappini C., 2007, A\&A, 462, 943. doi:10.1051/0004-6361:20065403
\bibitem[\protect\citeauthoryear{Churchwell et al.}{2009}]{Churchwell2009} Churchwell E., Babler B.~L., Meade M.~R., Whitney B.~A., Benjamin R., Indebetouw R., Cyganowski C., et al., 2009, PASP, 121, 213. doi:10.1086/597811
\bibitem[\protect\citeauthoryear{Cleveland}{1979}]{Cleveland1979}
Cleveland W.,
1979, Journal of the American Statistical Association, 1979, 74, 829-836
\bibitem[\protect\citeauthoryear{Crawford et al.}{2010}]{Cr10}
Crawford S. M., Still M., Schellart P. et al.,
2010, SPIE 7737, 773725
\bibitem[\protect\citeauthoryear{da Silva et al.}{2016}]{daSilva2016} da Silva R., Lemasle B., Bono G., Genovali K., McWilliam A., Cristallo S., Bergemann M., et al., 2016, A\&A, 586, A125. doi:10.1051/0004-6361/201527300
\bibitem[\protect\citeauthoryear{Davies et al}{2009}]{Davies2009}
Davies B., Origlia L., Kudritzki R.-P., et al.,
2009, ApJ, 696, 2014
\bibitem[\protect\citeauthoryear{Den Hartog et al.}{2003}]{DenHartog2003} Den Hartog E.~A., Lawler J.~E., Sneden C., Cowan J.~J., 2003, ApJS, 148, 543. doi:10.1086/376940
\bibitem[\protect\citeauthoryear{Donor et al}{2020}]{Donor2020}
Donor J., Frinchaboy P.~M., Cunha K.,
2020, AJ, 159, 199
\bibitem[\protect\citeauthoryear{Gaia Collaboration et al.}{2018}]{Gaia2018} Gaia Collaboration, Brown A.~G.~A., Vallenari A., Prusti T., de Bruijne J.~H.~J., Babusiaux C., Bailer-Jones C.~A.~L., et al., 2018, A\&A, 616, A1. doi:10.1051/0004-6361/201833051
\bibitem[\protect\citeauthoryear{Genovali et al.}{2013}]{Genovali2013} Genovali K., Lemasle B., Bono G., Romaniello M., Primas F., Fabrizio M., Buonanno R., et al., 2013, A\&A, 554, A132. doi:10.1051/0004-6361/201321650
\bibitem[\protect\citeauthoryear{Genovali et al}{2014}]{Genovali2014}
Genovali K., Lemasle B., Bono G., and 15 more
2014, A\&A, 566A, 37G
\bibitem[\protect\citeauthoryear{Gravity Collaboration}{2019}]{Gravity2019}
Gravity Collaboration; Abuter, R.;  Amorim, A.;  Baub\"ock, M.;
Berger, J. P.;  Bonnet, H.; Brandner, W.;  Cl\'enet, Y.;  et al.
2019, A\&A, 625, L10;
\bibitem[\protect\citeauthoryear{Ho et al.}{2017}]{Ho2017} Ho I.-T., Seibert M., Meidt S.~E., Kudritzki R.-P., Kobayashi C., Groves B.~A., Kewley L.~J., et al., 2017, ApJ, 846, 39. doi:10.3847/1538-4357/aa8460
\bibitem[\protect\citeauthoryear{Ho et al.}{2018}]{Ho2018} Ho I.-T., Meidt S.~E., Kudritzki R.-P., Groves B.~A., Seibert M., Madore B.~F., Schinnerer E., et al., 2018, A\&A, 618, A64. doi:10.1051/0004-6361/201833262
\bibitem[\protect\citeauthoryear{Hou}{2021}]{Hou2021} Hou L.~G., 2021, FrASS, 8, 103. doi:10.3389/fspas.2021.671670
\bibitem[\protect\citeauthoryear{Inno et al.}{2019}]{Inno2019} Inno L., Urbaneja M.~A., Matsunaga N., Bono G., Nonino M., Debattista V.~P., Sormani M.~C., et al., 2019, MNRAS, 482, 83. doi:10.1093/mnras/sty2661
\bibitem[\protect\citeauthoryear{Kniazev, Gvaramadze \& Berdnikov}{2016}]{KGB16}
   Kniazev A., Gvaramadze V. \& Berdnikov L., 2016, MNRAS, 459, 3068
\bibitem[\protect\citeauthoryear{Kniazev et al.}{2019}]{Kniazev2019}
  Kniazev A. Yu., Usenko I. A., Kovtyukh V. V., Berdnikov, L. N.
  2019, AstBu 74, 208
\bibitem[\protect\citeauthoryear{Korotin et al}{2014}]{Korotin2014}
Korotin S.~A., Andrievsky S.~M., Luck R.~E., L{\'e}pine J.~R.~D., Maciel W.~J., Kovtyukh V.~V.,
2014, MNRAS, 444, 3301
\bibitem[\protect\citeauthoryear{Kovtyukh et al}{2005}]{Kovtyukh2005}
Kovtyukh V.V., Wallerstein G., Andrievsky S.M.
2005, PASP, 117, 1173
\bibitem[\protect\citeauthoryear{Kovtyukh}{2007}]{Kov2007}
Kovtyukh V.~V., 2007, MNRAS, 378, 617
\bibitem[\protect\citeauthoryear{Kovtyukh et al.}{2016}]{Kovtyukh2016} Kovtyukh V., Lemasle B., Chekhonadskikh F., Bono G., Matsunaga N., Yushchenko A., Anderson R.~I., et al., 2016, MNRAS, 460, 2077. doi:10.1093/mnras/stw1113
\bibitem[\protect\citeauthoryear{Kovtyukh et al.}{2019}]{Kovtyukh2019} Kovtyukh V.~V., Andrievsky S.~M., Martin R.~P., Korotin S.~A., Lepine J.~R.~D., Maciel W.~J., Keir L.~E., et al., 2019, MNRAS, 489, 2254. doi:10.1093/mnras/stz2316
\bibitem[\protect\citeauthoryear{Kreckel et al.}{2019}]{Kreckel2019} Kreckel K., Ho I.-T., Blanc G.~A., Groves B., Santoro F., Schinnerer E., Bigiel F., et al., 2019, ApJ, 887, 80. doi:10.3847/1538-4357/ab5115
\bibitem[\protect\citeauthoryear{Kubryk, Prantzos, \& Athanassoula}{2015}]{Kubryk2015} Kubryk M., Prantzos N., Athanassoula E., 2015, A\&A, 580, A127. doi:10.1051/0004-6361/201424599
\bibitem[\protect\citeauthoryear{Kupka et al.}{1999}]{Kupka1999}
Kupka F., Piskunov N.E., Ryabchikova T.A., Stempels H. C., Weiss W. W., 1999, A\&AS, 138, 119
\bibitem[\protect\citeauthoryear{Lemasle et al.}{2007}]{Lemasle2007} Lemasle B., Fran{\c{c}}ois P., Bono G., Mottini M., Primas F., Romaniello M., 2007, A\&A, 467, 283. doi:10.1051/0004-6361:20066375
\bibitem[\protect\citeauthoryear{Lemasle et al.}{2008}]{Lemasle2008}
Lemasle B., Fran\c cois P.,  Piersimoni A.,  Pedicelli S.,  Bono G., Laney C. D., Primas F.,  Romaniello M.,
2008, A\&A, 490, 613
\bibitem[\protect\citeauthoryear{Lemasle et al}{2013}]{Lemasle2013}
Lemasle B., Fran\c cois P., Genovali K., et al.
2013, A\&A, 558, 31
\bibitem[\protect\citeauthoryear{Lemasle et al.}{2015}]{Lemasle2015}
   Lemasle B., Kovtyukh V., Bono G. et al.  2015, A\&A 579, A47
\bibitem[\protect\citeauthoryear{Li et al}{2013}]{Li2013}
Li Y., Bresolin F., Kennicutt R.~C., Jr., 
2013, ApJ, 766, 17
\bibitem[\protect\citeauthoryear{Li et al.}{2015}]{Li2015} Li C., Wang E., Lin L., Bershady M.~A., Bundy K., Tremonti C.~A., Xiao T., et al., 2015, ApJ, 804, 125. doi:10.1088/0004-637X/804/2/125
 \bibitem[\protect\citeauthoryear{{Luck}, {Gieren}, {Andrievsky}, {Kovtyukh},
   {Fouqu{\'e}}, {Pont} \& {Kienzle}}{{Luck} et~al.}{2003}]{Luck2003}
   {Luck} R.~E.,  {Gieren} W.~P., {Andrievsky} S.~M., {Kovtyukh} V.~V.,
   {Fouqu{\'e}} P.,  {Pont} F., {Kienzle} F., 2003, \aap, 401, 939
 \bibitem[\protect\citeauthoryear{Luck et al}{2006}]{Luck2006}
   Luck R.~E., Kovtyukh V.~V., Andrievsky S.~M.
   2006, AJ, 132, 902
\bibitem[\protect\citeauthoryear{Luck}{2018}]{Luck2018}
  Luck R. E., 2018, AJ 156, 171L 
\bibitem[\protect\citeauthoryear{Magrini et al.}{2017}]{Magrini2017} Magrini L., Randich S., Kordopatis G., Prantzos N., Romano D., Chieffi A., Limongi M., et al., 2017, A\&A, 603, A2. doi:10.1051/0004-6361/201630294
\bibitem[\protect\citeauthoryear{Mainzer et al.}{2011}]{Mainzer2011} Mainzer A., Bauer J., Grav T., Masiero J., Cutri R.~M., Dailey J., Eisenhardt P., et al., 2011, ApJ, 731, 53. doi:10.1088/0004-637X/731/1/53
\bibitem[\protect\citeauthoryear{Matteucci \& Francois}{1989}]{Matteucci1989} Matteucci F., Francois P., 1989, MNRAS, 239, 885. doi:10.1093/mnras/239.3.885
\bibitem[\protect\citeauthoryear{Martin et al.}{2015}]{Martin2015} Martin R.~P., Andrievsky S.~M., Kovtyukh V.~V., Korotin S.~A., Yegorova I.~A., Saviane I., 2015, MNRAS, 449, 4071. doi:10.1093/mnras/stv590
\bibitem[\protect\citeauthoryear{McWilliam, Wallerstein, \& Mottini}{2013}]{McWilliam2013} McWilliam A., Wallerstein G., Mottini M., 2013, ApJ, 778, 149. doi:10.1088/0004-637X/778/2/149
\bibitem[\protect\citeauthoryear{Minchev et al}{2014}]{Minchev2014}
Minchev, I., Chiappini, C., \& Martig, M. 2014, A\&A, 572, A92 
\bibitem[\protect\citeauthoryear{Moll{\'a} et al.}{2019}]{Molla2019} Moll{\'a} M., Wekesa S., Cavichia O., D{\'\i}az {\'A}. I., Gibson B.~K., Rosales-Ortega F.~F., Ascasibar Y., et al., 2019, MNRAS, 490, 665. doi:10.1093/mnras/stz2537
\bibitem[\protect\citeauthoryear{O'Donoghue et al.}{2006}]{Dono2006}
O'Donoghue D., Buckley D. A. H., Balona L. A. et al.,
2006, MNRAS, 372, 151
\bibitem[\protect\citeauthoryear{Origlia et al.}{2013}]{Origlia2013} Origlia L., Oliva E., Maiolino R., Mucciarelli A., Baffa C., Biliotti V., Bruno P., et al., 2013, A\&A, 560, A46. doi:10.1051/0004-6361/201322586
\bibitem[\protect\citeauthoryear{Pedicelli et al.}{2010}]{Pedicelli2010} Pedicelli S., Lemasle B., Groenewegen M., Romaniello M., Bono G., Laney C.~D., Fran{\c{c}}ois P., et al., 2010, A\&A, 518, A11. doi:10.1051/0004-6361/201014262
\bibitem[\protect\citeauthoryear{Proxauf et al.}{2018}]{Proxauf2018} Proxauf B., da Silva R., Kovtyukh V.~V., Bono G., Inno L., Lemasle B., Pritchard J., et al., 2018, A\&A, 616, A82. doi:10.1051/0004-6361/201833087
\bibitem[\protect\citeauthoryear{Reid et al.}{2019}]{Reid2019} Reid M. J.,  Menten K. M.,  Brunthaler A.,  Zheng X. W.,  Dame T. M., Xu Y.,  Li J., Sakai N. et al., 2019, ApJ 885, 131
\bibitem[\protect\citeauthoryear{S{\'a}nchez et al}{2015}]{Sanchez2015}
S{\'a}nchez S.~F., Galbany L., P{\'e}rez E. et al.,
2015, A\&A, 573, A105
\bibitem[\protect\citeauthoryear{S{\'a}nchez}{2020}]{Sanchez2020} S{\'a}nchez S.~F., 2020, ARA\&A, 58, 99. doi:10.1146/annurev-astro-012120-013326
\bibitem[\protect\citeauthoryear{S{\'a}nchez-Menguiano et al}{2016}]{Sanchez-Menguiano2016}
S{\'a}nchez-Menguiano L., S{\'a}nchez S.~F., Kawata D., et al.,
2016, ApJL, 830, L40
\bibitem[\protect\citeauthoryear{S{\'a}nchez-Menguiano et al.}{2018}]{Sanchez-Menguiano2018} S{\'a}nchez-Menguiano L., S{\'a}nchez S.~F., P{\'e}rez I., Ruiz-Lara T., Galbany L., Anderson J.~P., Kr{\"u}hler T., et al., 2018, A\&A, 609, A119. doi:10.1051/0004-6361/201731486
\bibitem[\protect\citeauthoryear{Schaye et al.}{2015}]{Schaye2015} Schaye J., Crain R.~A., Bower R.~G., Furlong M., Schaller M., Theuns T., Dalla Vecchia C., et al., 2015, MNRAS, 446, 521. doi:10.1093/mnras/stu2058
\bibitem[\protect\citeauthoryear{Skowron et al.}{2019}]{Skowron2019}
Skowron D. M.,  Skowron J., Mr{\'o}z P.,  Udalski A.,  Pietrukowicz P., Soszy{\'n}ski I., Szyma{\'n}ski M. K.,  Poleski R., Koz{\l}owski S.,  Ulaczyk K.,  Rybicki K.,  Iwanek P., Wrona M.,  Gromadzki M., 2019, AcA, 69, 305
\bibitem[\protect\citeauthoryear{Solar, Tissera, \& Hernandez-Jimenez}{2020}]{Solar2020} Solar M., Tissera P.~B., Hernandez-Jimenez J.~A., 2020, MNRAS, 491, 4894. doi:10.1093/mnras/stz2853
\bibitem[\protect\citeauthoryear{Spitoni \& Matteucci}{2011}]{Spitoni2011} Spitoni E., Matteucci F., 2011, A\&A, 531, A72. doi:10.1051/0004-6361/201015749
 \bibitem[\protect\citeauthoryear{Spitoni et al.}{2019}]{Spitoni2019}
   Spitoni E., Cescutti G., Minchev I., Matteucci F., Silva Aguirre V., Martig M., Bono G., Chiappini C.,
   2019, A\&A, 628A, 38     
\bibitem[\protect\citeauthoryear{{Twarog}, {Ashman} \& {Anthony-Twarog}}{{Twarog} et~al.}{1997}]{Twarog1997}
{Twarog} B.~A., {Ashman} K.~M., {Anthony-Twarog} B.~J., 1997, \aj, 114, 2556
\bibitem[\protect\citeauthoryear{Vasilyev et al.}{2019}]{Vasilyev2019} Vasilyev V., Amarsi A.~M., Ludwig H.-G., Lemasle B., 2019, A\&A, 624, A85. doi:10.1051/0004-6361/201935067
\bibitem[\protect\citeauthoryear{Vogt et al}{2017}]{Vogt2017}
Vogt F.~P.~A., P{\'e}rez E., Dopita M.~A., Verdes-Montenegro L., Borthakur S.,
2017, A\&A, 601, A61
\bibitem[\protect\citeauthoryear{Wang et al.}{2018}]{Wang2018} Wang S., Chen X., de Grijs R., Deng L., 2018, ApJ, 852, 78. doi:10.3847/1538-4357/aa9d99
\bibitem[\protect\citeauthoryear{Wenger et al.}{2019}]{Wenger2019} Wenger T.~V., Balser D.~S., Anderson L.~D., Bania T.~M., 2019, ApJ, 887, 114. doi:10.3847/1538-4357/ab53d3
\bibitem[\protect\citeauthoryear{Zinchenko et al.}{2016}]{Zinchenko2016} Zinchenko I.~A., Pilyugin L.~S., Grebel E.~K., S{\'a}nchez S.~F., V{\'\i}lchez J.~M., 2016, MNRAS, 462, 2715. doi:10.1093/mnras/stw1857
\end{thebibliography}
\end{document}